\documentclass[sn-nature]{sn-jnl}% Style for submissions to Nature Portfolio journals

\usepackage{graphicx}%
\usepackage{multirow}%
\usepackage{amsmath,amssymb,amsfonts}%
\usepackage{amsthm}%
\usepackage{mathrsfs}%
\usepackage[title]{appendix}%
\usepackage{xcolor}%
\usepackage{textcomp}%
\usepackage{manyfoot}%
\usepackage{booktabs}%
\usepackage{tabularx} % Add this package
\usepackage{algorithm}%
\usepackage{algorithmicx}%
\usepackage{algpseudocode}%
\usepackage{listings}%
\usepackage{tikz}
\usetikzlibrary{positioning, shapes.multipart, arrows.meta}
\usepackage[edges]{forest}
\usepackage{float}
\usepackage{hyperref}

\theoremstyle{thmstyleone}%
\theoremstyle{thmstyletwo}%

\theoremstyle{thmstylethree}%

\usepackage{geometry}
\begin{document}

\title[Article Title]{SAM vs BET: A Comparative Study for Brain Extraction and Segmentation of Magnetic Resonance Images using Deep Learning}


\author[1,2]{\fnm{Sovesh} \sur{Mohapatra}}

\author[2]{\fnm{Advait} \sur{Gosai}}

\author*[1,3,4]{\fnm{Gottfried} \sur{Schlaug}}\email{gschlaug@umass.edu}

\affil[1]{\orgdiv{Institute of Applied Life Sciences}, \orgname{University of Massachusetts}, \orgaddress{\city{Amherst}, \postcode{01003}, \state{MA}}}

\affil[2]{\orgdiv{Manning College of Information and Computer Sciences}, \orgname{University of Massachusetts}, \orgaddress{\city{Amherst}, \postcode{01003}, \state{MA}}}

\affil[3]{\orgdiv{Department of Biomedical Engineering}, \orgname{University of Massachusetts}, \orgaddress{\city{Amherst}, \postcode{01003}, \state{MA}}}

\affil[4]{\orgdiv{Department of Neurology}, \orgname{Baystate Medical Center, and UMass Chan Medical School - Baystate Campus}, \orgaddress{\city{Springfield}, \postcode{01107}, \state{MA}}}


\abstract{Brain extraction is a critical preprocessing step in various neuroimaging studies, particularly enabling accurate separation of brain from non-brain tissue and segmentation of relevant within-brain tissue compartments and structures using Magnetic Resonance Imaging (MRI) data. FSL's Brain Extraction Tool (BET), although considered the current gold standard for automatic brain extraction, presents limitations and can lead to errors such as over-extraction in brains with lesions affecting the outer parts of the brain, inaccurate differentiation between brain tissue and surrounding meninges, and susceptibility to image quality issues. Recent advances in computer vision research have led to the development of the Segment Anything Model (SAM) by Meta AI, which has demonstrated remarkable potential in zero-shot segmentation of objects in real-world scenarios. In the current paper, we present a comparative analysis of brain extraction techniques comparing SAM with a widely used and current gold standard technique called BET on a variety of brain scans with varying image qualities, MR sequences, and brain lesions affecting different brain regions. We find that SAM outperforms BET based on average Dice coefficient, IoU and accuracy metrics, particularly in cases where image quality is compromised by signal inhomogeneities, non-isotropic voxel resolutions, or the presence of brain lesions that are located near (or involve) the outer regions of the brain and the meninges. In addition, SAM has also unsurpassed segmentation properties allowing a fine grain separation of different issue compartments and different brain structures. These results suggest that SAM has the potential to emerge as a more accurate, robust and versatile tool for a broad range of brain extraction and segmentation applications.}




\maketitle 

 \section{Main}\label{intro}

Brain extraction from MRI or other imaging techniques is an important preprocessing step to advance various further image segmentation tasks \cite{despotovic2015mri}. This process involves isolating  brain tissue from non-brain tissues, such as the skull, scalp, and meninges, while preserving brain tissue that might have been altered due to various types of lesions or injuries, enabling accurate and effective subsequent processing \cite{ashburner2003computer,somasundaram2010fully,thompson1997detection}. Automatic and user independent brain extraction is challenged when brain tissue or non-brain tissue have similar or inhomogeneous signal intensity, leading to confusion around what constitutes brain tissue. The accurate removal of non-brain tissue enables automatic brain segmentation models to achieve significantly higher accuracy, and the reduction in information being processed enhances computational efficiency, making this an indispensable component in the preprocessing pipeline of brain imaging studies \cite{van2008brain,fein2006statistical}.

The current gold standard in automatic brain extraction is a tool offered by the FMRIB Software Library (FSL). However, this Brain Extraction Tool (BET) has several limitations, such as over-extraction leading to loss of brain matter, inaccurate extraction resulting in residual skull or non-brain tissue surrounding the brain to be falsely identified and included as brain tissue, and sensitivity to image quality, all of which complicates its application on clinical images \cite{zhuang2006skull, klein2010evaluation, boesen2004quantitative}. Consequently, manual intervention is often required to correct over and under-cropping errors, especially when dealing with scans exhibiting atypical features such as lesions, injuries, or uncommon image sequences \cite{smith2002fast, jenkinson2012fsl, battaglini2008enhanced}. Furthermore, BET's restricted scope confines its utility to general brain extraction, without providing the capability to isolate or segment specific regions or structures of the brain to be studied individually. Some areas of interest might include the corpus callosum, the grey and white matter, subcortical structures like the basal ganglia, vascular structures within the brain, and any kind of brain lesions. 

As a result, numerous deep learning-based models have been developed for brain extraction and other region specific extraction related tasks. Despite this, there is still a significant demand for enhancing the accuracy of their results \cite{pei2022general, fatima2020state}. These models heavily rely on training data and have exhibited a lack of robustness \cite{thakur2020brain}. As a consequence, BET remains the most widely used whole brain extraction tool, largely due to its extraction speed and familiarity within the field.  

The Segment Anything Model by Meta AI (SAM) has emerged as a state-of-the-art tool for image segmentation of any kind, being trained over a vast dataset of 1 billion masks and 11 million images \cite{kirillov2023segment}. SAM has already shown remarkable potential in accurately segmenting objects in real-world scenarios; its extensive training and zero-shot learning allow it to respond appropriately to any prompt at inference time \cite{he2022masked,xian2018zero}. Its demonstrated speed, accuracy and versatility position it as a promising tool for overcoming the limitations of BET and other deep learning-based models for whole brain extraction, as well as segmenting isolated brain regions or structures.

To explore the potential of SAM in neuroimaging research, we conducted a comparative analysis of brain extraction techniques and tested the capability of SAM to segment structures within the brain. Our study evaluated both SAM and BET on a diverse range of brain scans for brain extraction, which included varying image qualities and brain challenges caused by lesions and injuries. Additionally, we highlight various additional benefits that SAM could offer to neuroimaging researchers over existing techniques in rapidly segmenting within-brain structures of translational research and clinically applied relevance.

\section{Methodology}\label{methods}

\subsection{Selection of MRI data} \label{select mri}

We examined 45 anonymized MR brain images curated from various databases (ATLAS \cite{liew2018large}, white matter hyperintensities (WMH) Challenge Segmentation \cite{kuijf2019standardized}, and our anonymized and de-identified in-house dataset of neuroimages). To encompass a wide range of factors, including the various types of MRI sequences utilized, quality of the images, presence of any brain abnormalities, and use of both normalized and non-normalized data, we sampled five scans from nine customized brain image categories each (as presented in Section \ref{results}).

\subsection{Preprocessing of MRI data} \label{preprocessing}

To enable more accurate and reliable comparisons between the results obtained from BET and SAM, we resampled all the MRI data into MNI152 space \cite{evans2012brain}. This approach was intended to ensure all data is aligned to a standardized reference template, which is essential when performing group-level analyses and comparing results across different outputs.

\subsection{Brain Extraction using BET} \label{bet}

BET, a command-line tool, offers the convenience of using various parameters to enhance brain extraction in difficult or unusual cases, resulting in a customizable and semi-automated brain extraction process. The primary parameter which is used is the fractional intensity threshold ($f$) that determines the intensity level to distinguish between brain and non-brain tissue. By adjusting this value, we optimized the brain boundary estimation for the various data sets used in this paper. Notably, BET is able to perform brain extraction directly on 3D data, without requiring conversion into 2D slices. 

In certain instances, we found that none of the $f$ values effectively isolated the brain without either including a significant amount of non-brain tissue or excessively truncating the brain. In such cases, we reported the extraction results obtained using the default value ($f=0.5$).

\begin{figure}[h]
    \centering
    \includegraphics[width=0.8\textwidth]{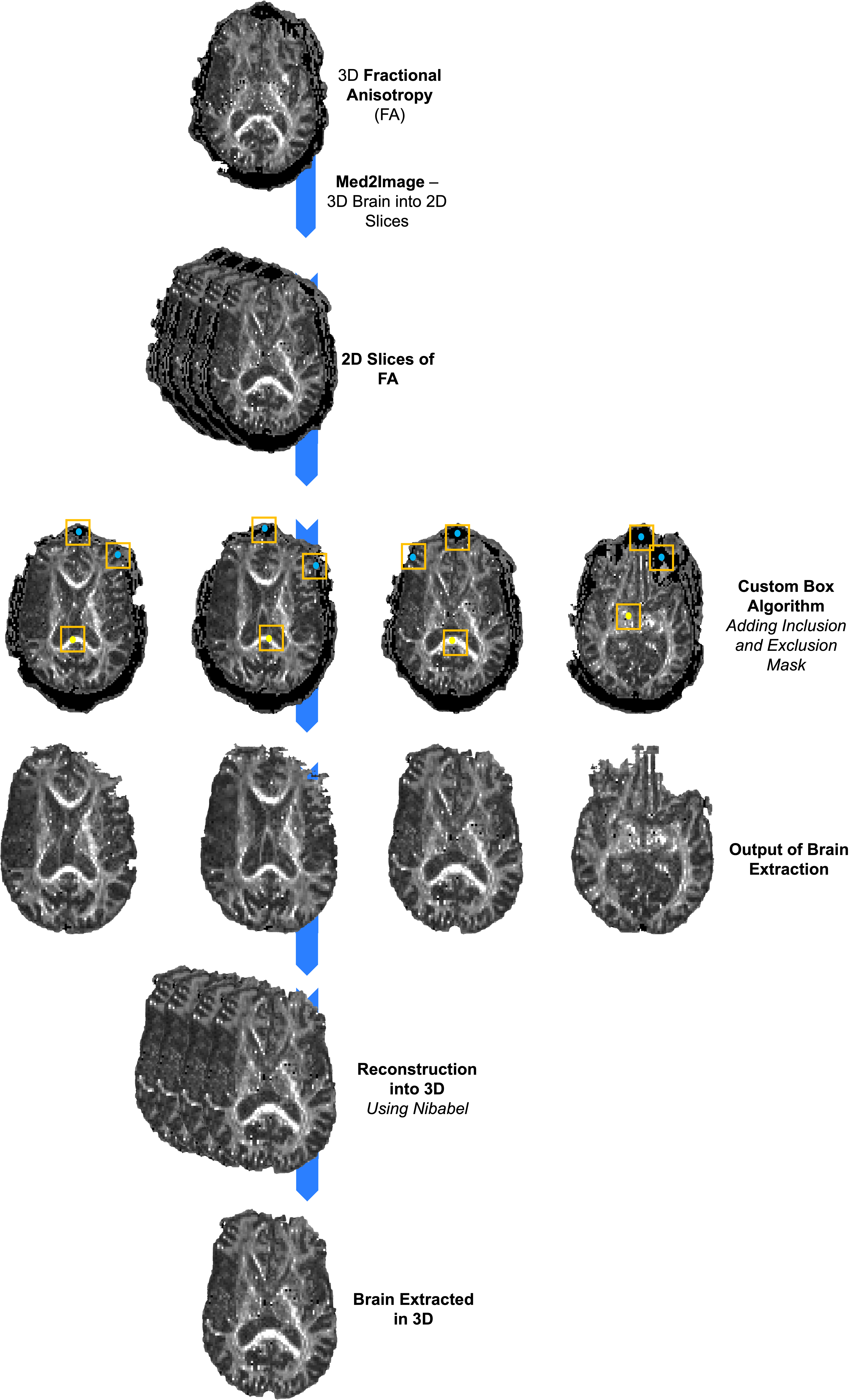}
    \caption{Comprehensive SAM workflow – visualizing the end-to-end process with custom box algorithm, highlighting blue dots (exclusion mask) and yellow dots (inclusion mask)}
    \label{fig-f}
\end{figure}

\subsection{Brain Extraction using SAM} \label{SAM}

SAM facilitated the customization of the brain extraction process by enabling us to utilize various parameters, such as intensity threshold and spatial location, to selectively retain or remove different regions of the image. As a result, we were able to achieve a fully automated and highly customizable extraction process, not only for the brain but also for specific regions within the brain, such as the corpus callosum, subcortical structures (e.g., basal ganglia), vascular structures, and any other distinct brain structures of interest such as brain lesions (refer to \ref{4.2}). 

In our case, we initially transformed each MRI scan into a set of 2D slices. Subsequently, we implemented a custom bounding box algorithm to accurately position markers, allowing us to effectively include a particular region or structure within the brain for extraction while simultaneously excluding unwanted regions based on their corresponding and neighboring voxel intensities. These inclusion and exclusion markers, alongside the 2D image slices, served as the set of inputs enabling precise mask generation from SAM.

\subsubsection{3D MRI scan preprocessing and reconstruction}

SAM requires 2D images as input, which we obtained by converting the processed 3D MRI data into various 2D slices using the med2image Python tool. The predicted slices were then converted into the required brain extracted data using the nibabel Python library. The end-to-end workflow can be found in Figure \ref{fig-f}.

\subsection{Evaluation metrics for brain extraction}

We have evaluated our results over five different metrics: Dice coefficient, Jaccard Index (referred to as IoU), Accuracy, Precision, and Recall, in order to understand how the outputs from BET and SAM have performed against the standard MNI152 1mm Brain images (as shown in the figure in SI). These images are used as ground truths for an extracted brain since they comprise of only relevant brain tissue. Detailed definitions of each of the chosen evaluation metrics and their relevance can be found in the supplemental information.

\begin{table}[b]
    \centering
    \caption{Comparison of the evaluation metrics for 9 customized categories of brain images. The source of the data is mentioned in parentheses next to the type of MRI sequence. The numbers in bold indicate the value of the metric, and the superior model for the specific brain image type. (Acc = Accuracy, Prec = Precision)}
    \label{tab:1}
    \addtolength{\tabcolsep}{-3pt}
    \begin{tabular}{ccccccccccc}
        \hline \\[-1.5ex]
        \multicolumn{1}{c}{} & \multicolumn{5}{c}{\textbf{BET}} & \multicolumn{5}{c}{\textbf{SAM}} \\[1ex]
        \textbf{Category} & Dice & IoU & Acc & Recall & Prec & Dice & IoU & Acc & Recall & Prec \\[1ex]
        \hline \\[-1.5ex]
        T1-weighted images \\ from Acute Stroke Phase (ATLAS \cite{liew2018large}) & 0.942 & 0.891 & 0.919 & 0.945 & 0.940 & \textbf{0.956} & \textbf{0.918} & \textbf{0.936} & \textbf{0.958} & \textbf{0.957}\\[1ex]
        T2-weighted images \\ from Chronic  Stroke Phase (in-house) & 0.966 & 0.936 & 0.951 & \textbf{0.985} & 0.957 & \textbf{0.969} & \textbf{0.940} & \textbf{0.955} & 0.971 &\textbf{0.967}  \\[1ex]
        FLAIR images \\ from Chronic Stroke Phase (in-house) & 0.949 & 0.907 & 0.928 & 0.945 & 0.954 & \textbf{0.955} & \textbf{0.915} & \textbf{0.935} & \textbf{0.954} &\textbf{0.957} \\[1ex]
        Normalized T2-weighted images \\ from Chronic Stroke Phase (in-house) & 0.919 & 0.854 & 0.860 & \textbf{0.990} & 0.862 & \textbf{0.977} & \textbf{0.955} & \textbf{0.965} & \textbf{0.990} &\textbf{0.965}  \\[1ex]
        Normalized FLAIR images \\ from Chronic Stroke Phase (in-house) & 0.853 & 0.744 & 0.743 & \textbf{0.951} & 0.792 & \textbf{0.962} & \textbf{0.927} & \textbf{0.941} & \textbf{0.951} &\textbf{0.962} \\[1ex]
        FLAIR images \\ from WMH Challenge Set \cite{kuijf2019standardized}* & \textbf{0.937} & \textbf{0.883} & \textbf{0.904} & \textbf{0.941} & 0.934 & 0.928 & 0.868 & 0.893 & 0.905 & \textbf{0.953} \\[1ex]
        3DT1-weighted images \\ from WMH Challenge Set \cite{kuijf2019standardized} & 0.628 & 0.518 & 0.642 & 0.549 & \textbf{0.956} & \textbf{0.914} & \textbf{0.842} & \textbf{0.871} & \textbf{0.944} & 0.888 \\[1ex]
        T1-weighted images \\ from WMH Challenge Set \cite{kuijf2019standardized} & 0.898 & 0.815 & 0.857 & 0.863 & \textbf{0.936} & \textbf{0.928} & \textbf{0.867} & \textbf{0.898} & \textbf{0.923} & 0.934 \\[1ex]
        Fractional Anisotropy (FA) images \\ from Chronic Stroke Phase (in-house) & 0.461 & 0.305 & 0.480 & 0.306 & \textbf{0.988} & \textbf{0.916} & \textbf{0.845} & \textbf{0.880} & \textbf{0.893} & 0.941 \\[1ex]
        \hline
    \end{tabular}
    *Results discussed in Section \ref{sec3.1}. Refer to the SI document for more detail. 
\end{table}

\section{Results}\label{results}

\subsection{Evaluation of Extractions by BET and SAM}\label{sec3.1}

In our evaluation of brain extractions performed by BET and SAM, we compared their performance across five key metrics against ground truth. Our analysis demonstrated that SAM outperformed BET in the majority of cases, as evidenced by the evaluation metrics and accompanying figures from various imaging modalities. The supplemental information contains the visual comparisons and raw scores for each of the 45 scans used to generate our results.

The results presented in Table \ref{tab:1} indicate that both SAM and BET exhibit high scores across all five metrics when processing high-quality T2-weighted images without lesions, or images that have undergone spatial normalization to a standard space. While normalization does not lead to a significant improvement in BET’s performance, SAM’s metrics exhibit a minor improvement compared to its non-normalized samples.

For our in-house datasets, BET’s metrics experience a decline when dealing with lower quality FLAIR images (both normalized and non-normalized). This is also observed for Acute Stroke Phase images from the ATLAS dataset \cite{liew2018large}, potentially due to lesion size and positioning. In contrast, the custom bounding box algorithm implemented with SAM enables high performance across most image qualities and cases of brain abnormalities, as it allows dynamic inclusion and exclusion of specific smaller regions based on their intensities. The most evident examples of this include our in-house Fractional Anisotropy (FA) images and 3DT1 weighted images from the WMH challenge set. For instance, in the WMH 3DT1 dataset, our SAM based algorithm achieved a Dice coefficient of 0.914, an IoU of 0.842, an accuracy of 0.871, and a recall of 0.944, This is in contrast to BET's performance, which yielded a Dice coefficient of 0.628, an IoU of 0.518, an accuracy of 0.642, and a recall of 0.306. 

For both of these datasets, we observe a notably higher BET precision, even though there is poor performance on other metrics. This can be attributed to the excessive cropping by BET, which inadvertently removes large portions of actual brain tissue. Consequently, the substantially smaller segmented region does not contain any out-of-brain tissue, leading to a minimal number of false positives which results in deceptively high precision value (as shown in Table \ref{tab:1}). 

SAM also presents room for improvement. This is particularly seen for the FLAIR images from the WMH dataset \cite{kuijf2019standardized} where SAM slightly underperforms BET, as highly specific exclusion markers near the skull region result in occasional over-extraction (as shown in Figure I of the supplemental information).

\begin{figure}[h]
    \centering
    \includegraphics[width=1\textwidth]{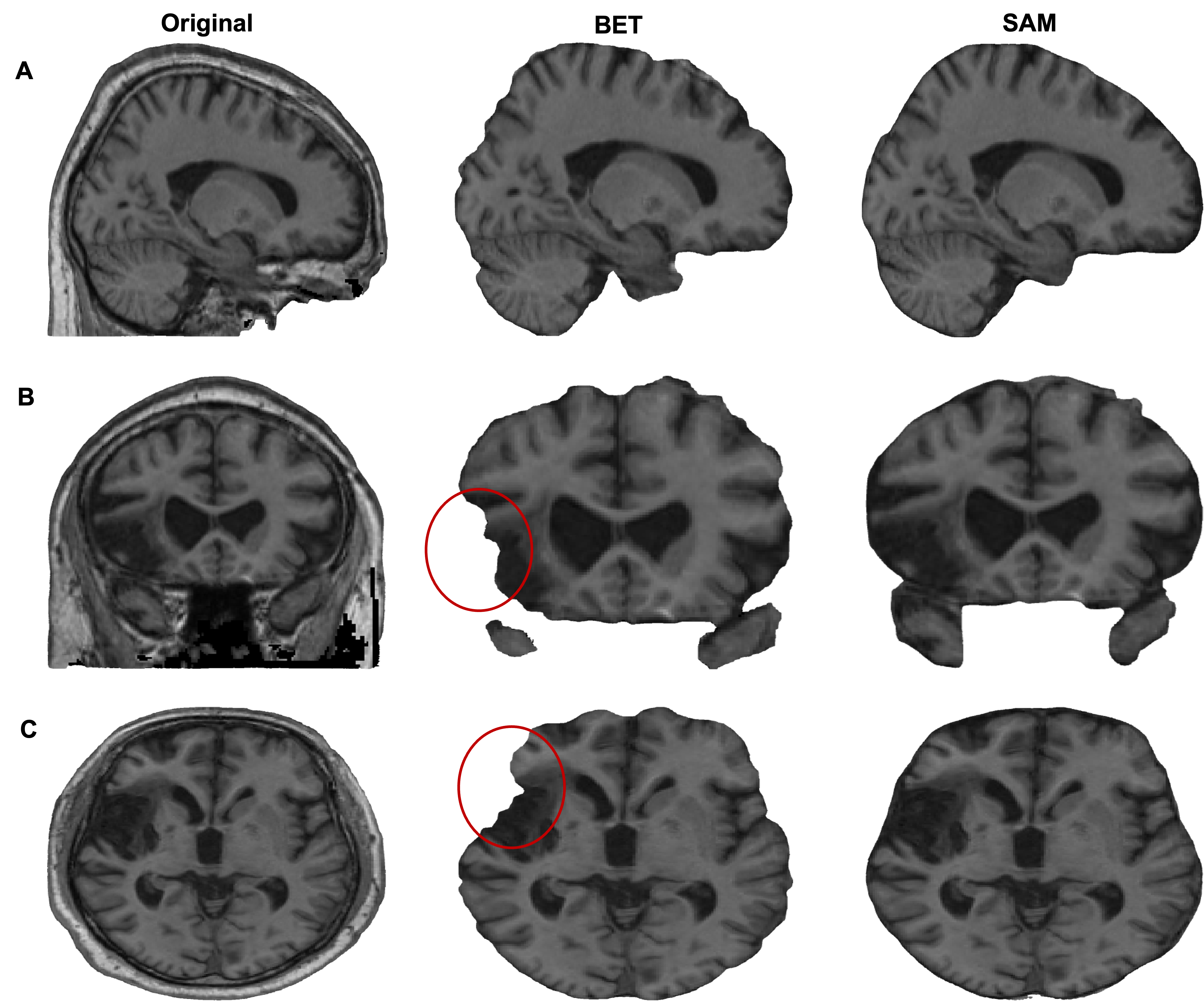}
    \caption{Comparison of original MRI images and extraction outputs by BET and SAM for \textbf{A.} sagittal, \textbf{B.} coronal, and \textbf{C.} axial anatomical planes. BET extraction results in sagittal, coronal, and axial planes, respectively. SAM extraction results in sagittal, coronal, and axial planes, respectively. The data is from ATLAS v2.0 dataset \cite{liew2018large}.}
    \label{fig2}
\end{figure}

\subsection{Visual Comparison}\label{vis}

Our findings from the figures are consistent with those presented in Table \ref{tab:1}, whereby BET performed highly consistent and desirable extractions in cases involving high-quality images and normalized data, however, SAM consistently outperformed BET across most other variations. For instance, Figure \ref{fig2} depicts the T1 scan from the acute stroke phase in the ATLAS v2.0 dataset, along with its corresponding BET output and SAM output across three different planes: sagittal, coronal, and axial. Here, we notice that BET's performance is compromised since the scan contained lesions proximal to the brain's outer contour or in regions that are challenging to distinguish from the surrounding meninges, despite good image quality.  

\begin{figure}[b]
    \centering
    \includegraphics[width=1\textwidth]{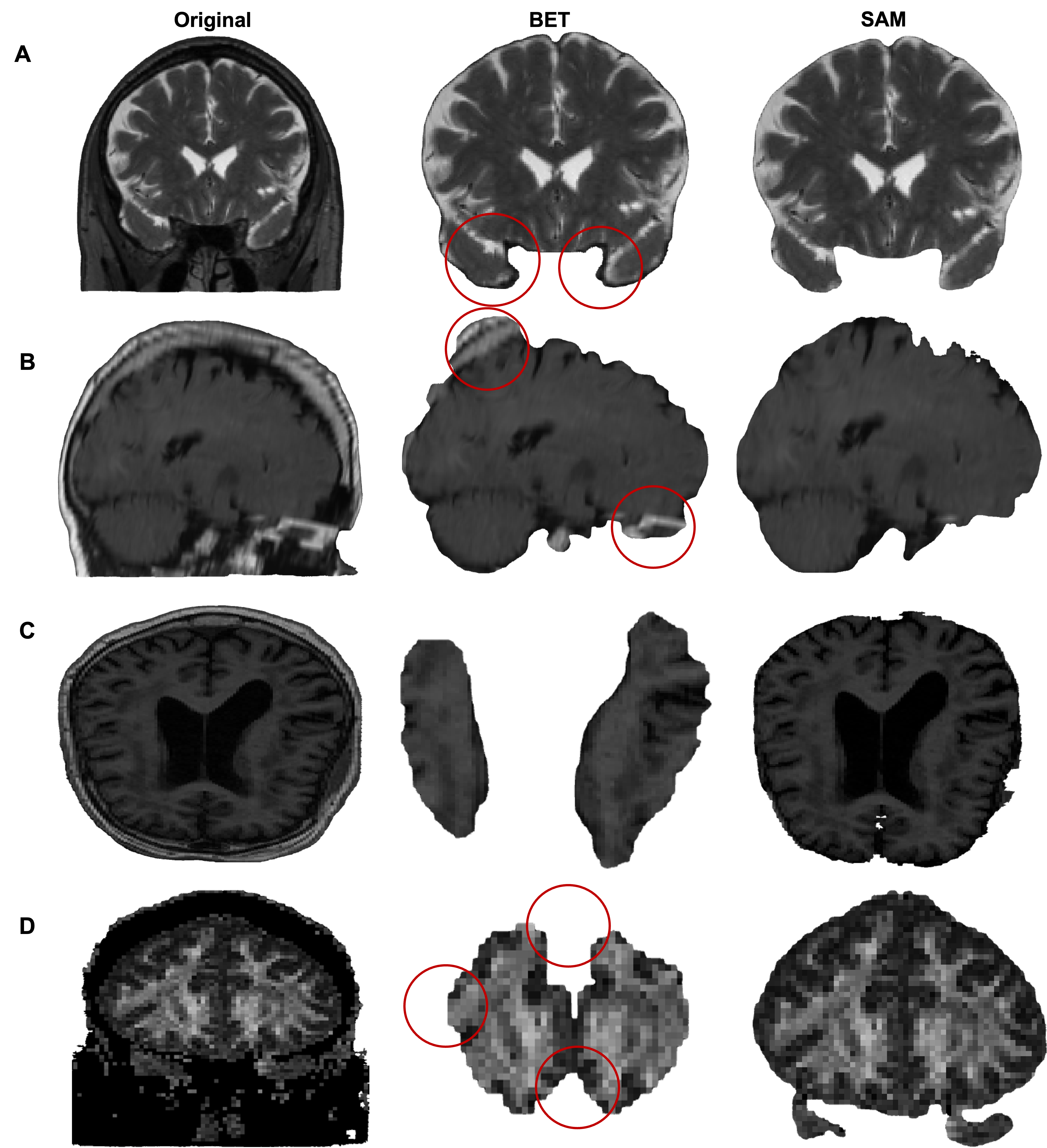}
    \caption{Comparison of original MRI images and extraction outputs by BET and SAM for \textbf{A.} T2 scan from a chronic stroke patient in the coronal plane, \textbf{B.} FLAIR scan from a chronic stroke patient in the sagittal plane, \textbf{C.} 3DT1 scan from a patient having WMH in the coronal plane, and \textbf{D.} FA scan from a patient having chronic stroke in the coronal plane. Data for \textbf{A., B. and D.} are from in-house dataset, and \textbf{C.} is from WMH challenge dataset \cite{kuijf2019standardized}.}
    \label{fig3}
\end{figure}

In Figure \ref{fig3}, we can observe one of the three planes of the original brain, along with its corresponding BET and SAM outputs for a random sample of the MRI modalities we studied. We observe that both BET and SAM produced comparable brain extractions for high resolution normalized T2-weighted images (\ref{fig3}A), and for the in-house FLAIR image (\ref{fig3}B) where BET leaves part of the skull, whereas SAM performs minor over-cropping. However, the limitations of BET related to lesion proximity to the outer parts of the brain and image quality are clearly evident in \ref{fig3}C and \ref{fig3}D, where the extracted brains are severely over-cropped, rendering them unsuitable for any future segmentation or subsequent analysis. This highlights a disadvantage of using BET in such cases, whereas SAM's consistently accurate extractions offer a more reliable option.

\subsection{Applications of SAM segmenting particular brain structures and lesions}\label{4.2}

\begin{figure}[h]
    \centering
    \includegraphics[width=0.9\textwidth]{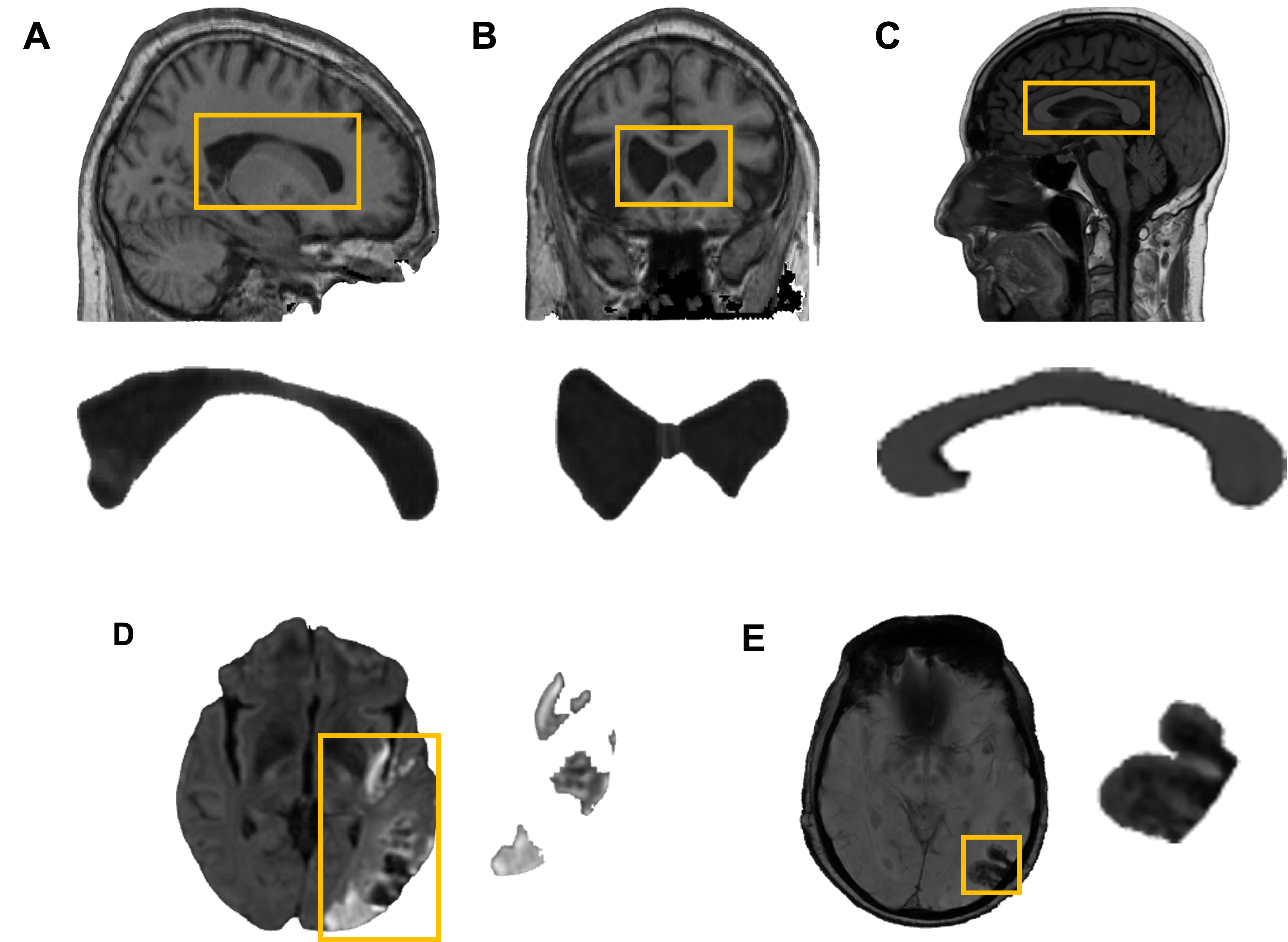}
    \caption{\textbf{A., B.} Ventricles extracted from a T1 scan from acute stroke phase. \textbf{C.} Corpus callosum extracted from a 3DT1 scan from chronic stroke phase. \textbf{D.} Lesions extracted from a DWI scan from chronic stroke phase. \textbf{E.} Microbleeds extracted from a SWAN scan from chronic stroke phase. Data in \textbf{A. and B.} are from ATLAS v2.0 \cite{liew2018large}, and \textbf{C., D., and E.} are from in-house dataset.}
    \label{fig4}
\end{figure}

Given its versatility, the potential applications of SAM in neuroimaging research are vast. This tool could be used for automating the various preprocessing pipelines required in imaging data for large scale studies like those investigating effects and changes in neurodegenerative diseases or brain development.  Additionally, its ability to accurately segment and extract brain structures, as well as identifying specific regions of interest, could accelerate a lot of the studies involving structure-function relationships as well as quantitative lesion volume and lesion load determination.  This could also  contribute and enhance precision medicine and personalized medicine in neurology and psychiatry. As demonstrated in Figure \ref{fig4}, we show examples of how SAM could also be utilized for precise segmentation of brain regions and brain lesions demonstrating its utility in segmenting ischemic and hemorrhagic lesions, specific brain structures of interest such as the midsagittal corpus callosum size, and subcomponents of brain that might be affected in particular neurological disorders such as the size/volumes of ventricles or size/volume of basal ganglia structures.

\section{Discussion}

The results of our study provided valuable comparative insights into the effectiveness and various
potential applications and challenges of a new method of brain extraction, SAM, with BET, a gold-standard in the field of brain extraction. The primary objective of our study was to examine the performance of these tools in the context of brain extraction and provide an overview of how effective SAM is in segmenting the specific regions of interest. Our findings support the utility of these tools for specific tasks of image pre-processing as well as detailed within-image processing
and analysis.

In accordance with our research question, the study has shown that both the SAM and BET are capable of accurately extracting brain structures from a variety of brain types to a certain degree of error. In addition, the comparison of these tools has revealed that SAM has outperformed BET in terms of Dice coefficient, Jaccard Index and Accuracy for a majority of our diverse set of tested cases. SAM’s robustness has been highlighted based on both quantitaive and qualitative analysis, suggesting its potential to be a more suitable tool for brain extraction, segmentation, and identification of specific regions in neuroimaging research when paired with an appropriate inclusion/exclusion algorithm.

\subsection{Advantages and Limitations of SAM and BET}

One of the key advantages of SAM is its flexibility and adaptability to different types of data. Since the model has been trained on a huge number of images, it enables it to adapt to various imaging modalities and contacts, making it a versatile tool for a wide range of neuroimaging applications. Moreover, SAM’s impressive performance on all evaluation parameters attests to its efficacy in brain extraction. Its ability to identify specific brain regions, such as the corpus callosum, subcortical structures, vascular structures, and other specialized areas, makes it a highly versatile tool. Nonetheless, it is important to note that the conversion of 3D data to 2D slices and subsequent reconstruction, which is required for using SAM, represents a potential drawback and source of
error in neuroimaging analysis. 

BET’s strength lies in its simplicity, speed, and ease of use, with it being a reliable tool for brain extraction particularly when brains are in a spatially normalized space. However, our study validated that its extraction accuracy is quite inferior to that of SAM in certain scenarios, such as
when the image quality is suboptimal, has poor contrast, unnormalized, or when images contain brain lesions that are located near the outer portions of the brain. Moreover, unlike SAM, BET is unable to efficiently segment and extract specific brain sub-regions, which limits its application to primarily full brain extraction.

Finally, it is worth noting that processing the data using SAM requires high computational resources. The entire pipeline, from converting 3D data to running SAM on 2D slices and per- forming reconstruction, can take approximately 30 seconds, which is considerably longer than BET. 

\subsection{Future Work}

Future research in this area should concentrate on addressing the existing limitations of SAM, such as exploring various reconstruction techniques to optimize its performance for 3D image datasets. Section \ref{4.2} provides a demonstration of possible applications of SAM based segmentation algorithms in neuroimaging research outside of brain extraction which are vital to be explored in greater detail. These algorithms can be further refined to involve a variety of relevant parameters to improve the quality of inclusion, exclusion markers, or generate extremely precise bounding boxes for automatic mask extraction from SAM. To enhance the robustness and generalizability of SAM inherently across a broad range of contextual tasks, including fast and precise lesion segmentation, segmentation of clinically relevant structures within the brain, as well as lesion progression and therapy responses, fine-tuning could be performed on its baseline model. This would require a diverse and comprehensive training dataset that covers various neuroimaging modalities and pathologies.

\section{Conclusion}

Our findings provide evidence of the significant additional potential of SAM in comparison with BET in pre-processing as well as detailed within-brain image analysis, in particularly brain extraction from the skull and meninges even in challenging cases with lesions involving the outer parts of the brain, brain tissue segmentation allowing the differentiation of lesioned from non-lesioned tissue components, and rapid identification and delineation of specific brain regions of interest. Our evaluation metrics indicate that SAM consistently outperforms BET, demonstrating its superior performance across a broad range of imaging modalities and pathologies. Our ongoing research aims to further improve the effectiveness and generalizability of the algorithm across a wide array of neuroimages and to utilize SAM’s versatility for other brain structure segmentation and extraction tasks. We believe that advancements in the fields of image preprocessing and segmentation can lead to a deeper understanding of brain structure and function and can facilitate the development of more precise and effective diagnostic approaches, which has the potential to advance therapeutic options. Overall, our study highlights the value of cutting-edge deep learning techniques to advance the field of applied and translational neuroimaging.

\section*{Supplemental Information}
Here is the link to the supplementary document referenced in this paper: shorturl.at/bACZ9

\section*{Acknowledgement}

This research was supported by NIMH (Brain-Initiative) (7R01MH111874-05). GS also acknowledges support from NINDS (U01NS102353). SM acknowledges the support from IALS summer internship program (direct summer stipend support). Conflict of interest statement: GS is the co-founder of Brainify, LLC, and has ownership interest in Brainify, LLC. Brainify LLC financially supported the research conducted in this study, but had no influence on data processing or the derived results or the presentation or interpretation of the results.

\vfill
\pagebreak


\bibliography{sn-bibliography}% common bib file

\end{document}